\documentclass{jkas}

\def\year{2024} 
\def\volume{57} 
\def\issue{**} 
\def\beginpage{1} 
\def\received{February 19, 2024} 
\def\accepted{May 4, 2024} 
\def\published{June **, 2024} 
\date{Received \received; Accepted \accepted; Published \published}

\usepackage{flushend} 
\usepackage{xcolor}

\title{Impact of Postshock Turbulence on the Radio Spectrum of Radio Relic Shocks in Merging Clusters}

\author{Hyesung Kang}{0000-0002-4674-568}

\affil{Department of Earth Sciences, Pusan National University, Busan 46241, Korea; }

\def\corrauthor{H. Kang, \email{hskang@pusan.ac.kr}}

\def\runningauthor{Kang}
\def\runningtitle{Impact of Postshock Turbulence on Radio Relics}

\def\keywords{acceleration of particles --- cosmic rays ---
galaxies: clusters: general --- shock waves}

\def\abstracttext{
This study investigates the impact of magnetic turbulence on cosmic ray (CR) electrons through Fermi-II acceleration behind merger-driven shocks in the intracluster medium and examines how the ensuing synchrotron radio emission is influenced by the decay of magnetic energy through dissipation in the postshock region.
We adopt simplified models for the momentum diffusion coefficient, specifically considering transit-time-damping resonance with fast-mode waves and gyroresonance with Alfv\'en waves. 
Utilizing analytic solutions derived from diffusive shock acceleration theory, at the shock location, we introduce a CR spectrum  that is either shock-injected or shock-reaccelerated. We then track its temporal evolution along the Lagrangian fluid element in the time domain. 
The resulting CR spectra are mapped onto a spherical shell configuration to estimate the surface brightness profile of the model radio relics.
Turbulent acceleration proves to be a significant factor in delaying the aging of postshock CR electrons, while decaying magnetic fields have marginal impacts due to the dominance of inverse Compton cooling over synchrotron cooling. However, the decay of magnetic fields substantially reduces synchrotron radiation.  Consequently, the spatial distribution of the postshock magnetic fields affects the volume-integrated radio spectrum and its spectral index. We demonstrate that the Mach numbers estimated from the integrated spectral index tend to be higher than the actual shock Mach numbers, highlighting the necessity for accurate modeling of postshock magnetic turbulence in interpreting observations of radio relics.
}

\begin{document}
\jkashead 

\section{Introduction\label{s1}}

Giant radio relics found in the outskirts of galaxy clusters, such as the Sausage and Toothbrush relics, are thought to result from shocks that occur following the passage of the dark matter (DM) core during major mergers \citep[e.g.,][]{vanweeren10,vanweeren16,ha2018}.
They are weak quasi-perpendicular shocks with low Mach numbers ($M_s \lesssim 3$) formed in the weakly magnetized intracluster medium (ICM) \citep[e.g.,][]{kang12,kang2016,kang2017b}. 
Diffuse radio emissions originate from cosmic ray (CR) electrons with the Lorentz factor $\gamma\sim 10^3-10^4$, gyrating in microgauss-level magnetic fields. These electrons are believed to be accelerated via diffusive shock acceleration (DSA) \citep[see][for reviews]{brunetti2014,vanweeren2019}.
Alternative scenarios such as adiabatic compression by shocks \citep{ensslin01,ensslin02}, reacceleration of fossil CR electrons by shocks \citep{kang12,pinzke2013}, and reacceleration by postshock turbulence \citep{fujita2015,kang2017a} have been considered as well.

The DSA theory predicts that the energy spectrum of CR particles, accelerated through the Fermi first-order (Fermi-I) process, follows a power-law distribution, $f_{\rm sh}\propto p^{-q}$, where $q=4M_s^2/( M_s^2-1)$ \citep{bell1978,drury1983}. 
Consequently, this leads to a synchrotron radio spectrum, $j_{\nu} \propto \nu^{-\alpha_{\rm sh}}$ with the so-called ``injection spectral index'', $\alpha_{\rm sh}=(q-3)/2$, immediately behind the shock.
As a result, the Mach numbers of {\it radio relic shocks} can be estimated using the relation \citep[e.g.,][]{kang15}: 
\begin{equation}
M_{\rm rad,sh}=\left(\frac{3+2\alpha_{\rm sh}}{2\alpha_{\rm sh}-1}\right)^{1/2}.
\label{Mrad1}
\end{equation}
Alternatively, one can determine the Mach numbers by observing the steepening of the volume-integrated spectrum, $J_{\nu} \propto \nu^{-\alpha_{\rm int}}$, toward the so-called ``integrated spectral index", $\alpha_{\rm int}=\alpha_{\rm sh}+0.5$, at high frequencies. 
This steepening is attributed to synchrotron and inverse-Compton (IC) losses in the postshock region with a constant magnetic field strength, leading to the following relation \citep[e.g.,][]{kang2017b} :
\begin{equation}
M_{\rm rad,int}=\left(\frac{ \alpha_{\rm int}+1}{\alpha_{\rm int}-1}\right)^{1/2}. 
\label{Mrad2}
\end{equation}
However, the transition of the power-law index from $\alpha_{\rm sh}$ to $\alpha_{\rm int}$ takes place gradually over the broad frequency range of $\sim 0.1-10$~GHz, depending on the shock age and postshock magnetic field strength. 
Furthermore, the volume-integrated emission spectrum could deviate from the simple DSA power-law in the case of the evolving shock dynamics and nonuniform magnetic field strength in the postshock regions, as suggested by \citet{kang15}. 
Thus, the estimation of $M_{\rm rad,int}$ of observed radio relics tend to be higher than $M_{\rm rad,sh}$ \citep[e.g.,][]{hoang2018}.

On the other hand, Mach numbers inferred from X-ray observations, $M_{\rm X}$, are sometimes found to be smaller than $M_{\rm rad}$, i.e., $M_{\rm X}\lesssim M_{\rm rad}$ \citep[e.g.,][]{akamatsu13,vanweeren2019}. 
This discrepancy is recognized as an unsolved challenge in understanding the origin of radio relics. 
\citet{wittor2021} compiled values of $M_{\rm rad}$ and $M_{\rm X}$ for observed radio relics available in the literature,
confirming the Mach number discrepancy (refer to their Figure 7).
By employing cosmological structure formation simulations, the authors confirmed the prevailing notion that radio flux is dominated by contributions from high Mach number shocks among the ensemble associated with the particular relic, whereas X-ray emission predominantly originates from low Mach number shocks \citep[see also][]{hong2015, roh2019,botteon2020,DF2021}.
Additionally, several potential solutions have been suggested to address this puzzle. These include the reacceleration of preexisting fossil CR electrons with a flat spectrum \citep[e.g.,][]{pinzke2013,kang2016, kang2017b} and acceleration by multiple shocks with different Mach numbers formed in the turbulent ICM \citep[e.g.,][]{inchingolo2022}.

As clusters form through numerous merging episodes of smaller subclusters, the gas flows within the ICM inherently become turbulent \citep{miniati2015,poter2015,vazza2017}.
During active mergers, the ICM turbulence becomes transonic, and the largest turbulent eddies ($L\sim 100-500$~kpc) undergo decay into smaller ones. 
This process cascades into magnetohydrodynamic (MHD) turbulence and further down to kinetic turbulence through plasma instabilities, as comprehensively reviewed by \citet{brunetti2014}.
Additionally, vorticity generated behind curved ICM shocks is known to produce MHD turbulence and amplify magnetic fields in the postshock region \citep{ryu2008}.

On the other hand, numerical simulations of non-driven, decaying MHD turbulence indicate that turbulent energy dissipates within one eddy turnover time, $t_{\rm dec} \sim \lambda_d/v_{\rm turb}$, where $\lambda_d$ represents the largest driving scale, and $v_{\rm turb}$ is the mean turbulent velocity \citep[e.g.][]{maclow1998,maclow1999,cho2003}.
Consequently, behind typical merger shocks, the estimated turbulent decay timescale is approximately $t_{\rm dec} \sim L/u_2 \sim {(100~\rm kpc)}/({10^3~\rm km~s^{-1}}) \sim 0.1$~Gyr, where $L$ is the largest eddy size of the induced turbulence and $u_2$ is the characteristic postshock flow speed\footnote{Throughout the paper, the subscript `2' is used for the postshock quantities.}.

Moreover, the interaction of preexisting turbulence with shock waves can induce corrugation of the shock front, thereby enhancing postshock turbulence on plasma kinetic scales through processes such as shock compression and turbulent dynamo mechanisms \citep{guo2015, trotta2023}.
Hybrid kinetic simulations of similar setups also indicate that postshock magnetic fluctuations exhibit a Kolmogorov spectrum and undergo substantial decay downstream due to dissipation \citep{nakanotani2022}.
Although these studies examined the plasma processes and wave-particle interactions on kinetic scales in a low beta ($\beta= P_B/P_g\sim 1$) plasma relevant for interplanetary shocks, we expect the same processes to operate similarly in the postshock region of ICM shocks formed in high beta ($\beta\sim 100$) plasma as well.

The amplification of postshock magnetic fields and the subsequent decay of MHD turbulence affects the radio spectrum of relic shocks. First, CR electrons can be further energized via Fermi second-order (Fermi-II) acceleration primarily through the interaction with the compressible fast mode waves via the transit-time-damping (TTD) resonance \citep{brunetti2007,brunetti2014}, and Alfv\'en waves via gyroresonance \citep{brunetti2004,fujita2015}.
Additionally, the synchrotron emission scales with the magnetic field strength as $j_{\nu}\propto B^{(q-1)/2}$, typically with $q\sim 4.0-5.0$, so the decay of magnetic fields $B$ significantly reduces synchrotron radiation emission.

\begin{figure*}[t]
\centering
\includegraphics[width=150mm]{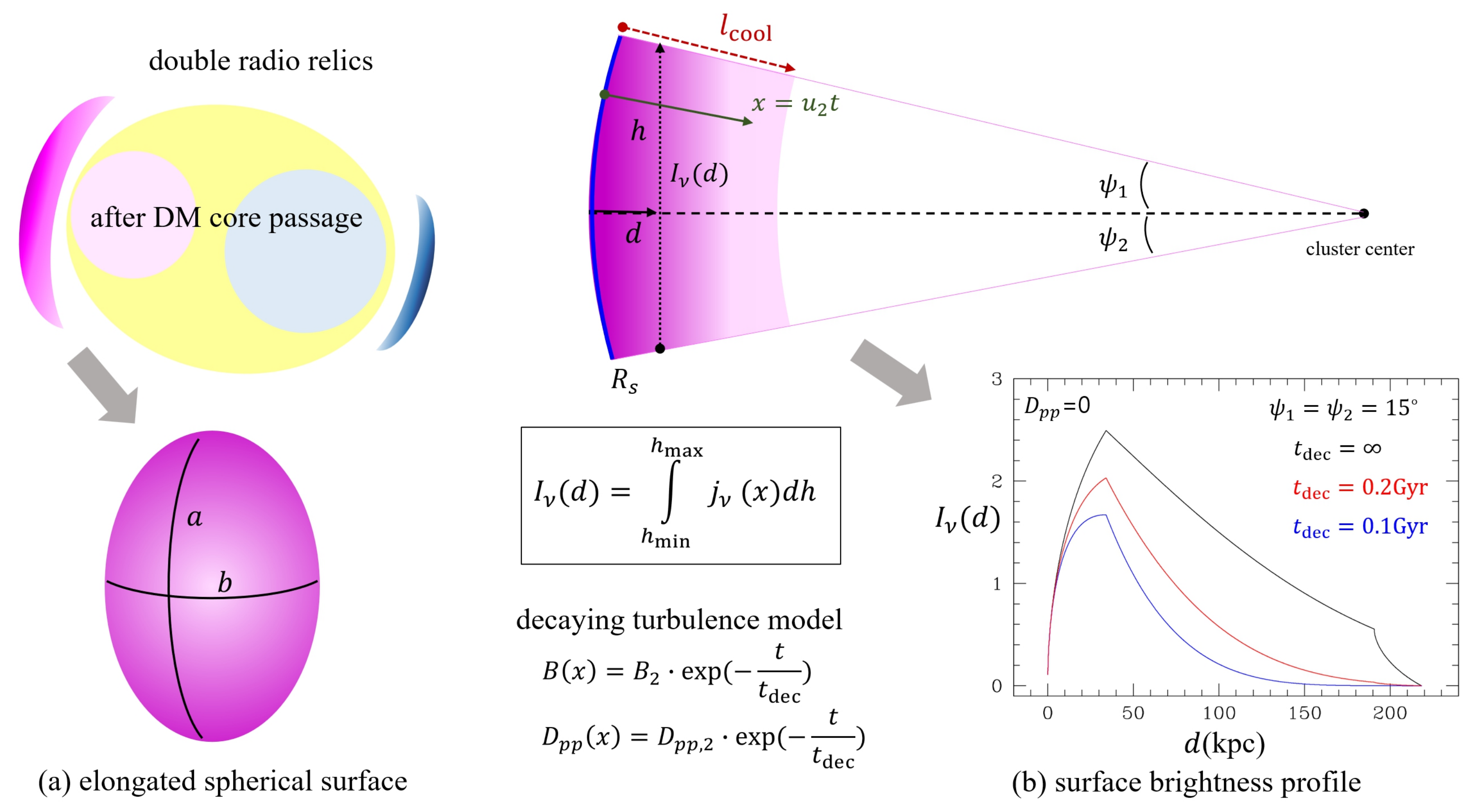}
\caption{
Schematic diagrams elucidate our model assumptions. (a) To model the surface of a radio relic, we employ a spherical, coconut-shell-shaped structure with an axial ratio of $a/b\gtrsim 1$ and a thickness of $l_{\rm cool}= u_2 t_{\rm cool}$. Here, $u_2$ and $t_{\rm cool}$ represent the advection speed and cooling timescale in the post-shock flow, respectively. Radio relics become prominent after the passage of the dark matter core during a major merger. (b) The surface brightness, $I_{\nu}(d)$, is estimated by integrating the volume emissivity, $j_{\nu}(x)$, with $x=u_2 t$, along a line of sight, where $d$ is the distance from the relic edge projected onto the sky plane. $I_{\nu}(d)$ depends on the CRe density, the magnetic field strength, $B(x)$, and the momentum diffusion coefficient, $D_{pp}(x)$, which decay with a timescale of $t_{\rm dec}$. Here, $B_2$ and $D_{\rm pp,2}$ are the immediate postshock values. The inset panel illustrates how the spatial profile of $I_{\nu}(d)$ depends on the decay timescale, $t_{\rm dec}$, of magnetic turbulence. 
Here, turbulent acceleration is ignored ($D_{pp}=0$), but synchrotron and inverse-Compton losses are included. 
The shell radius is $R_s=1$~Mpc, and the extension angles are $\psi_1=\psi_2=15^{\circ}$.
\label{f1}}
\end{figure*}

In this study, we explore the impact of turbulent acceleration (TA) on the evolution of the CR electron spectrum in the postshock flow, considering the decay of the magnetic fluctuations. 
The numerical procedure is outlined as follows:
\begin{enumerate}
\item We incorporate Fermi-II acceleration of CR electrons, employing simplified models for the momentum diffusion coefficient, $D_{pp}(p)$. This accounts for TTD resonance with fast mode waves and gyroresonance with Alfv\'en waves.
\item We track the time evolution of the CR electron population, $f(p,t)$, by following the Lagrangian fluid element through advection in the postshock region. This is accomplished by solving the Fokker-Planck equation in the time domain. In a one-dimensional (1D) planar shock configuration, the time integration can be transformed into the spatial profile of $f(p,x)$ through the relation $x=u_2 t$, where $u_2$ is a constant postshock speed and $t$ is the advection time since the shock passage.
\item The synchrotron emissivity, $j_{\nu}(t)$, is calculated, utilizing the information for $f(p,t)$ and $B(t)$.
\item  The surface brightness profile, $I_{\nu}(d)$, is estimated as a function of the distance $d$ from the relic edge projected onto the sky plane. This is obtained by adopting a coconut-shell-shaped spherical surface, as illustrated in Figure \ref{f1}.
\end{enumerate}

In the next section, we provide detailed explanations of the numerical methods and working models employed to simulate physical processes.
In Section \ref{s3}, we apply our approach to various examples. 
Specifically, we focus on scenarios involving the injection and the reacceleration of CR electrons by weak shocks with Mach numbers $2.3 \lesssim M\lesssim 3$. 
Additionally, we estimate the resulting radio emission spectra in an idealized setup.
A brief summary of our findings will be presented in Section \ref{s4}.

\section{Physical Models and Numerical Method \label{s2}}

Here, we consider merger-driven shocks that become radio-luminous subsequent to the DM core passage in a major binary merger, as depicted in Figure \ref{f1}(a) \citep{ha2018}. 
Although the shock surface evolves as a spherical shell expanding radially, we treat its dynamics as a planar shock with a constant postshock speed. 
This simplification is justified because the thickness of the postshock volume is on the order of $l_{\rm cool}\approx u_2 t_{\rm cool}\sim 0.1$~Mpc, which is much smaller than the shock radius, $R_s\sim 1-1.5$~Mpc. 
Furthermore, the cooing timescale, $t_{\rm cool}\sim 0.1$~Gyr, is shorter than the typical dynamical timescales of clusters, $t_{\rm dyn}\sim 1$~Gyr. 
In such a scenario, the time integration can be transformed into the spatial profile using the relation $x=u_2 t$.

\subsection{Postshock Magnetic Turbulence\label{s2.1}}

\begin{figure}[t]
\centering
\includegraphics[width=80mm]{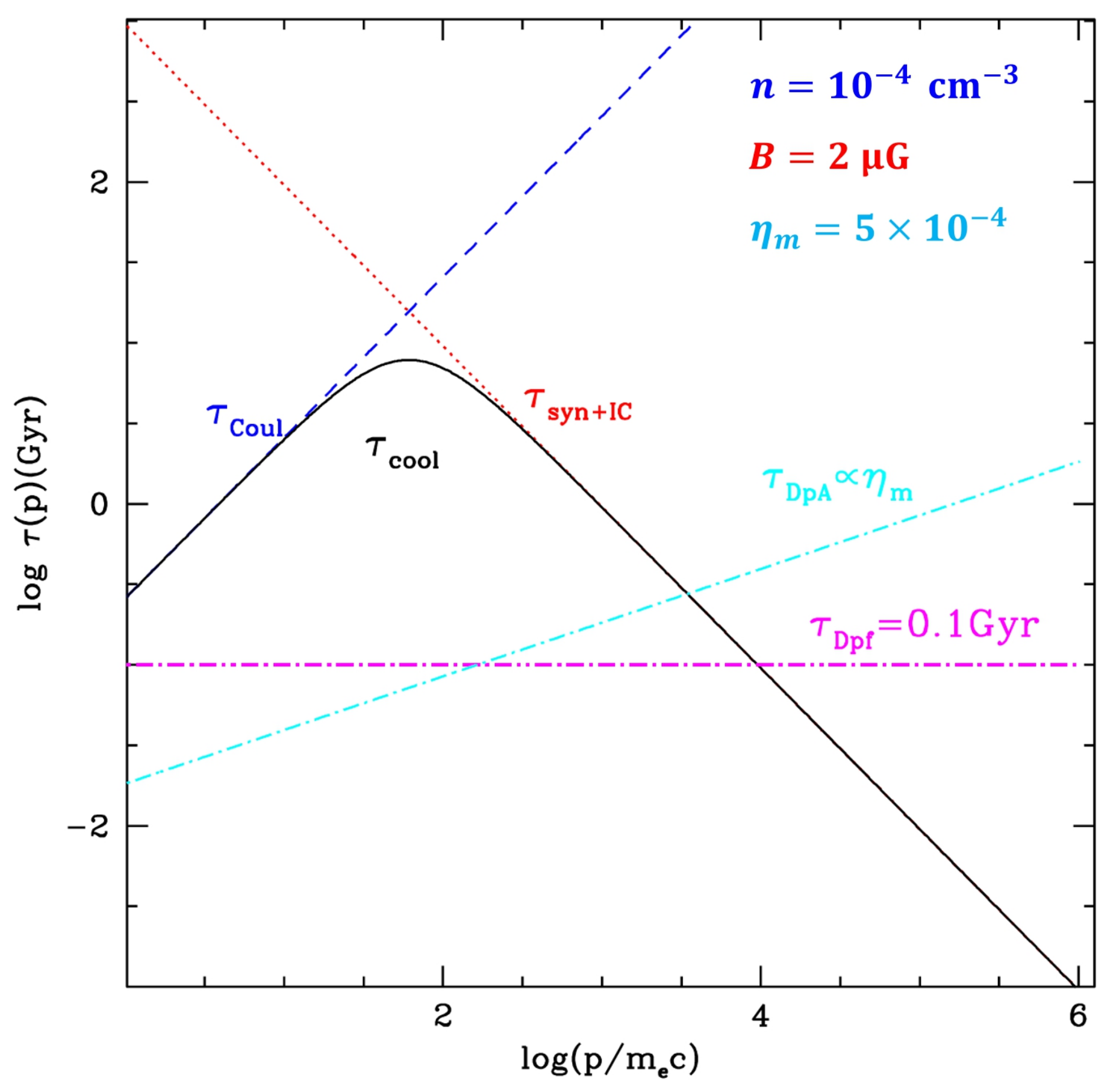}
\caption{Cooling timescales and TA timescales, all in units of $10^9$ years: $\tau_{\rm Coul}$ (blue) for Coulomb losses,
$\tau_{\rm Syn+IC}$ (red) for synchrotron and inverse Compton losses, $\tau_{\rm cool}$ (black) for the total losses,
$\tau_{\rm Dpf}$ (magenta) due to fast mode waves, and 
$\tau_{\rm DpA}$ (cyan) due to Alfv\'en mode waves. 
Representative cases are considered with the following parameters: 
gas density $n=10^{-4}~cm^{-3}$, magnetic field strength $B=2~\mu \rm G$, redshift $z_r=0.2$, and reduction factor, $\eta_m=5\times10^{-4}$.
\label{f2}}
\end{figure}

\begin{figure}[t]
\centering
\includegraphics[width=80mm]{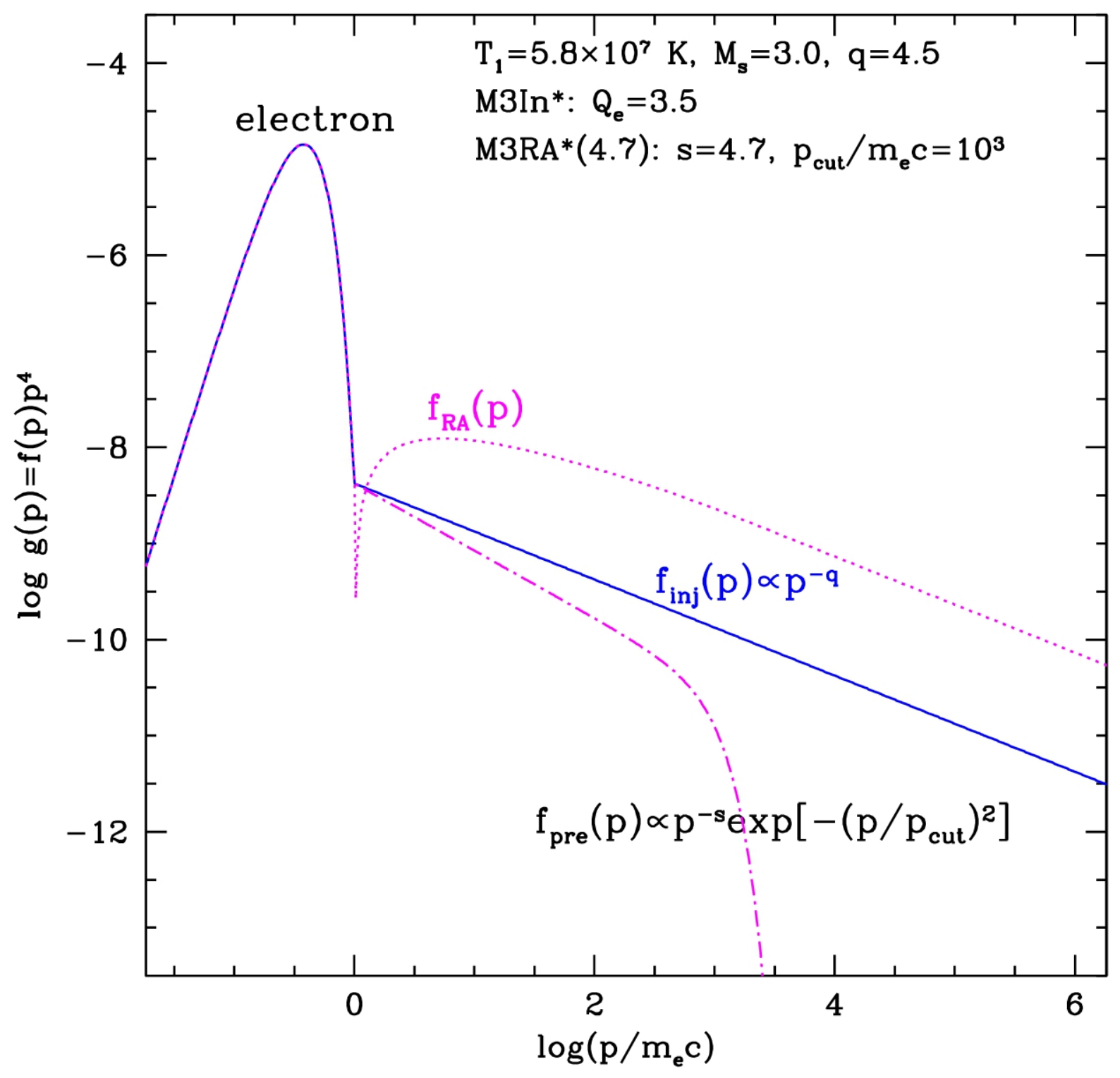}
\caption{The momentum distribution function, $g(p)=p^4f(p)$, is depicted in a $M_{\rm s}=3.0$ shock, based on the test-particle DSA model. The blue line represents the injected population, $f_{\rm inj}$. The magenta dotted-dashed line illustrates a power-law spectrum of the pre-existing fossil electron population, $f_{\rm pre}$, with a slope $s=4.7$ and a cutoff momentum $p_{\rm cut}/m_{\rm e}c=10^3$. 
The magenta dotted line displays the spectrum of the reaccelerated population, $f_{\rm RA}$.
Here, the amplitude of $f_{\rm pre}$ is the same as that of $f_{\rm inj}(p)$ at $p_{\rm min}=Q_e\cdot p_{\rm th}$, where $Q_e=3.5$.
In our calculations, $f_{\rm inj}$ and $f_{\rm RA}$ are deposited at the shock front ($t=0$) in the M3In* and M3RA* models, respectively.
\label{f3}}
\end{figure}

As outlined in the introduction, downstream of the shock front, CR electrons further acquire energy through TTD resonance with compressive fast-mode waves and gyroresonant scattering off Alfv\'en waves. These waves might be present in small-scale, kinetic magnetic fluctuations that are cascaded down from MHD-scale turbulence \citep{brunetti2014} or excited by plasma microinstabilities in the shock transition zone \citep{guo2015, trotta2023}. 
However, the microphysics governing the excitation and evolution of MHD/plasma waves and Fermi-II acceleration of CR electrons in the high beta ICM plasmas are quite complex and relatively underexplored \citep[e.g.][]{lazarian2012}.
This makes it hard to formulate accurate models for the momentum diffusion coefficient, $D_{pp}$.

The TA timescale due to the interaction with fast modes can be related with $D_{pp}$ as 
\begin{equation}
\label{Dpf}
\frac{D_{\rm pp,f}}{p^2} \approx \frac{4}{\tau_{pp}(p)},
\end{equation}
where, in general, $\tau_{pp}(p)$ depends on the nature and amplitude of magnetic fluctuations, $\delta B(x,t)$, in the flow.
As in many previous studies \cite[e.g.][]{kang2017b}, we take a practical approach in which a constant value, $\tau_{pp}=0.1$~Gyr is assumed since the detail properties of the postshock turbulence are not well constrained.
For instance, using cosmological structure formation simulations, \citet{miniati2015} found that in the ICM typically $t_{pp}\sim 0.1-1 $~Gyr due to enhanced turbulence during the active phase of major mergers.  

Based on the work of \citet{fujita2015}, we adopt $D_{pp}$ due to gyro-resonance with Alfv\'en waves as follows:
\begin{eqnarray}
\label{DpA}
\frac{D_{\rm pp,A}}{p^2} \sim  \frac{1}{9} (\frac{v_A^2}{D_{xx}}) \sim \frac{1}{3} (\frac{v_A}{c})(\frac{v_A}{l_{\rm mfp}}) \nonumber\\
\sim \frac{1}{3} (\frac{v_A}{c})(\frac{v_A}{l_{\rm mfp,c}}) \eta_m^{-1}(p/p_0)^{q_K-2}
\end{eqnarray}
where $v_A=B/\sqrt{4\pi \rho}$ is the Alfv\'en speed, $D_{xx}\sim c l_{\rm mfp}/3$ is the spatial diffusion coefficient, $\eta_m\sim 5\times 10^{-4}$ is a reduction factor for waves on small kinetic scales, and $p_0=10^{-3}m_ec$ is the reference momentum. 
The slope $q_K=5/3$ is adopted since Alfv\'en modes of decaying MHD turbulence are expected to have a Kolmogorov spectrum \citep[e.g.,][]{cho2003}.
As a result, $D_{\rm pp,A}/p^2\propto p^{-1/3}B^2 $, so TA becomes increasingly inefficient at higher momentum.
In addition, $D_{\rm pp,A}$ decreases as magnetic fluctuations decay in the postshock flow.
The Coulomb mean free path for thermal electrons can be estimated as
\begin{eqnarray}
l_{\rm mfp,c}\sim 174~{\rm kpc} (\frac{\ln \Lambda}{40})^{-1} (\frac{T}{10^8K})^2 (\frac{n}{10^{-4} {\rm cm^{-3}}})^{-1},
\end{eqnarray}
where $\ln \Lambda\sim 40$ is the Coulomb logarithm \citep{brunetti2007}.

Figure \ref{f2} shows the cooling timescales for Coulomb collisions, $\tau_{\rm Coul}$, and synchrotron plus IC losses, $\tau_{\rm sync+IC}$, for a representative set of parameters for the ICM, i.e., $n=10^{-4} {\rm cm^{-3}}$, $B=2~\mu G$, and the redshift, $z_r=0.2$.
For radio emitting CR electrons with the Lorentz factor, $\gamma\sim 10^3-10^4$, typical cooling timescales range $\tau_{\rm cool}\sim 0.1-1$~Gyr.
The figure also compares the TA timescales due to fast modes, $\tau_{\rm Dpf}$, and for Alfv\'en modes, $\tau_{\rm DpA}$.
For the set of representative parameters considered here, TA with $D_{\rm pp,A}$ is more efficient compared to radiative losses for $\gamma\lesssim 3\times 10^3$, whereas TA with $D_{\rm pp,f}$ is more efficient for $\gamma\lesssim 10^4$.

The full consideration of the evolution of postshock turbulence, including the vorticity generation behind a rippled shock front, additional injection of turbulence driven by continuous subclump mergers, decompression of the postshock flows, and kinetic wave-particle interactions, is beyond the scope of this study.
In anticipation of the dissipation of MHD turbulence energy, we employ an exponential function to model the decay of magnetic energy and the reduction of momentum diffusion: 
\begin{eqnarray}
B(t)= B_2 \cdot \exp(-t/t_{\rm dec}) \nonumber\\
D_{pp}(p,t) = D_{pp,2}(p) \cdot \exp(-t/t_{\rm dec}),
\label{Bdecay}
\end{eqnarray}
where $t_{\rm dec}=0.1$ or 0.2~Gyr is considered (see Table \ref{t1}).
Although the functional forms for the two quantities could differ with separate values of $t_{\rm dec}$, we opt for this simple model to reduce the number of free parameters in our modeling. 
In addition, we note that non-driven MHD turbulence is known to decay as a power law in time, i.e., $E_B\propto (1+ C_B t/t_{\rm dec})^{-\eta}$ with $\eta\sim 1$ and $C_B\sim 1$ \citep{maclow1998,cho2003}.
Within one eddy turnover time ($t_{\rm dec}$), the magnetic energy density decreases by a factor of $\sim 2.7^2$ in the exponential decay model given in equation (\ref{Bdecay}), and by a factor of $\sim 2$ in the power-law decline model.
We can justify our choice since our study primarily focuses on a qualitative examination of how turbulence decay influences postshock synchrotron emission.
Considering that $t_{\rm dec}$ is a not-so-well constrained, free parameter in our model, the quantitative interpretation of our results should be taken with caution.

\begin{table*}[t]
\caption{Model Parameters and Estimated Spectral Indices\label{t1}}
\centering
\begin{tabular}{lccccccccc}
\toprule
Model Name   & $D_{pp}$ & $t_{\rm dec}$(Myr) &  $f_{\rm inj}\propto p^{-q}$ & $(\alpha_{0.15}^{0.61})^a$ & $\alpha_{0.61}^{3.0}$ & $\alpha_{3.0}^{16}$ & $(M_{0.15}^{0.61})^b$ & $M_{0.61}^{3.0}$ & $M_{3.0}^{16}$\\
\midrule
M3InDp0      &  $D_{pp}=0$  & $\infty$ &  &1.15 &1.25 &1.25 & 3.80 & 3.01 & 2.97 \\
\midrule
M3InDp0(200) &  $D_{pp}=0$  &  200 &               &1.02&1.10 &1.18 &11.1&4.51 &3.52 \\
M3InDpf(200) &  $D_{\rm pp,f}$  &  200 &           & 1.09& 1.30  & 1.39 & 4.78 & 2.75 &2.49 \\
M3InDpA(200) &  $D_{\rm pp,A}$  &  200 &           & 1.18& 1.18 & 1.22& 4.45& 3.51 & 3.20\\
M3InDp0(100) &  $D_{pp}=0$  &  100    &            & 0.938 &1.03&1.12 & - & 8.68&4.22 \\
M3InDpf(100) &  $D_{\rm pp,f}$  &  100 &           & 0.985& 1.10 & 1.21 & - & 4.49&3.21 \\
M3InDpA(100) &  $D_{\rm pp,A}$  &  100 &           & 0.981& 1.06 & 1.14& -& 5.80& 3.86 \\
\midrule
Model Name   & $D_{pp}$  & $t_{\rm dec}$(Myr) & $f_{\rm pre}\propto p^{-s}$ & $\alpha_{0.15}^{0.61}$ & $\alpha_{0.61}^{3.0}$ & $\alpha_{3.0}^{16}$ & $M_{0.15}^{0.61}$& $M_{0.61}^{3.0}$ & $M_{3.0}^{16}$\\
\midrule
M3RADp0(4.3) &  $D_{pp}=0$  &  100    & $s=4.3$ & 0.938 &1.03 & 1.12& -& 8.68 &4.22\\
M3RADp0(4.7) &  $D_{pp}=0$  &  100    & $s=4.7$ & 0.938 &1.03 & 1.12& -& 8.68 &4.22\\
M3RADpf(4.3) &  $D_{\rm pp,f}$  &  100    & $s=4.3$ & 0.985  &1.10 & 1.21&-&4.49 &3.21\\
M3RADpf(4.7) &  $D_{\rm pp,f}$  &  100    & $s=4.7$ & 0.985 &1.10 & 1.21&-&4.49 &3.21\\
M3RADpA(4.3) &  $D_{\rm pp,A}$  &  100    & $s=4.3$ & 0.981 &1.06 & 1.14&-&5.80 &3.86\\
M3RADpA(4.7) &  $D_{\rm pp,A}$  &  100    & $s=4.7$ & 0.981 &1.06 & 1.14&-&5.80 &3.86\\
\bottomrule
\end{tabular}
\tabnote{
The model name consists of characters that represent the sonic Mach number, injection (In) or reacceleration (RA) cases, and the momentum diffusion models (Dp0, Dpf, and DpA).
For M3In*($t_{\rm dec}$) models, the number in the parenthesis is the decay time scale in units of Myr, while for M3RA*($s$) models, it is the power-law slope of the preexisting CR population.
The same set of models, M2.3*, for $M_s=2.3$ shocks are also considered. 
\\ $^{\rm a}$ The spectral index, $\alpha_{\nu_1}^{\nu_2}$, is estimated from the volume-integrated spectrum, $J_{\nu}$, between two frequencies, $\nu_1$ and $\nu_2$, where $\nu=0.15,~0.61,~3.0,~{\rm and}~16$~GHz. 
\\ $^{\rm b}$ The {\it integrated} Mach number, $M_{\nu_1}^{\nu_2}$, is estimated based on Equation (\ref{Mrad2}) using $\alpha_{\nu_1}^{\nu_2}$. Note that for $\alpha_{\nu_1}^{\nu_2}<1$, $M_{\nu_1}^{\nu_2}$ cannot be calculated.}
\end{table*}

\subsection{DSA CR Spectrum at the Shock Position\label{s2.2}}

We follow the time evolution of the CR distribution function, $f(p,t)$, in the Lagrangian fluid element that advects downstream with the constant postshock speed. So the spatial advection distance of the fluid element from the shock front is given as $x=u_2t$. At the shock position ($t=0$), the shock-injected spectrum, $f_{\rm inj}(p)$, or the shock-reaccelerated spectrum, $f_{\rm RA}(p)$, are assigned as the initial spectrum (see Figure \ref{f3}).

The spectrum of injected CR electrons is assumed to follow the DSA power-law for $p\ge p_{\rm min}$: 
\begin{equation}
f_{\rm inj}(p) \approx [{n_{\rm 2} \over \pi^{1.5}} p_{\rm th}^{-3} \exp(-Q_e^2)] \cdot \left(p \over p_{\rm min} \right) ^{-q},
\label{finj}
\end{equation}
where $n_2$ and $T_2$ are the postshock gas density and temperature, respectively \citep{kang2020}. 
In addition, $p_{\rm th}=\sqrt{2m_e k_B T_2}$, $p_{\rm min}= Q_e~ p_{\rm th}$ with the injection parameter $Q_e=3.5$.
Usual physical constants are used: $m_e$ for the electron mass, $c$ for the speed of light, and $k_B$ for the Boltzmann constant.

For the preshock population of CR electrons, we adopt a power-law spectrum with the slope $s$ for $p\ge p_{\rm min}$: 
\begin{equation}
f_{\rm pre}(p) = f_{\rm o} \cdot \left(p \over p_{\rm min} \right) ^{-s} \exp\left(-{p^2 \over p_{\rm cut}^2} \right),
\label{fpre}
\end{equation}
where $f_{\rm o}$ is the normalization factor and $p_{\rm cut}\approx 10^3m_e c$ is a cutoff momentum due to cooling.
The preexisting CR electrons may consist of fossil electrons injected by relativistic jets from radio galaxies or residual electrons accelerated in previous shock passages.
If these fossil electrons are accelerated by relativistic shocks contained in relativistic jets, the power-law slope could be $s\approx 4.3$ \citep{kirk2000}.
On the other hand, if they are accelerated by ICM shock with $M_s\approx 2.3-3$ in the cluster outskirts, $s\approx 4.5-4.9$ \citep{hong2014}.

The reaccelerated population at the shock can be calculated by the following integration:
\begin{equation}
f_{\rm RA}(p)= q \cdot p^{-q} \int_{p_{\rm min}}^p p^{\prime q-1} f_{\rm pre} (p^\prime) dp^\prime 
\label{freacc}
\end{equation}
\citep{drury1983,kang2011}.
Except in the case of $q=s$, $f_{\rm RA}(p)\propto p^{-r}$ with $r=\min (q,s)$,
meaning $f_{\rm RA}(p)$ adopts the harder spectrum between $p^{-q}$ and $p^{-s}$.

\begin{figure*}[t]
\centering
\includegraphics[width=150mm]{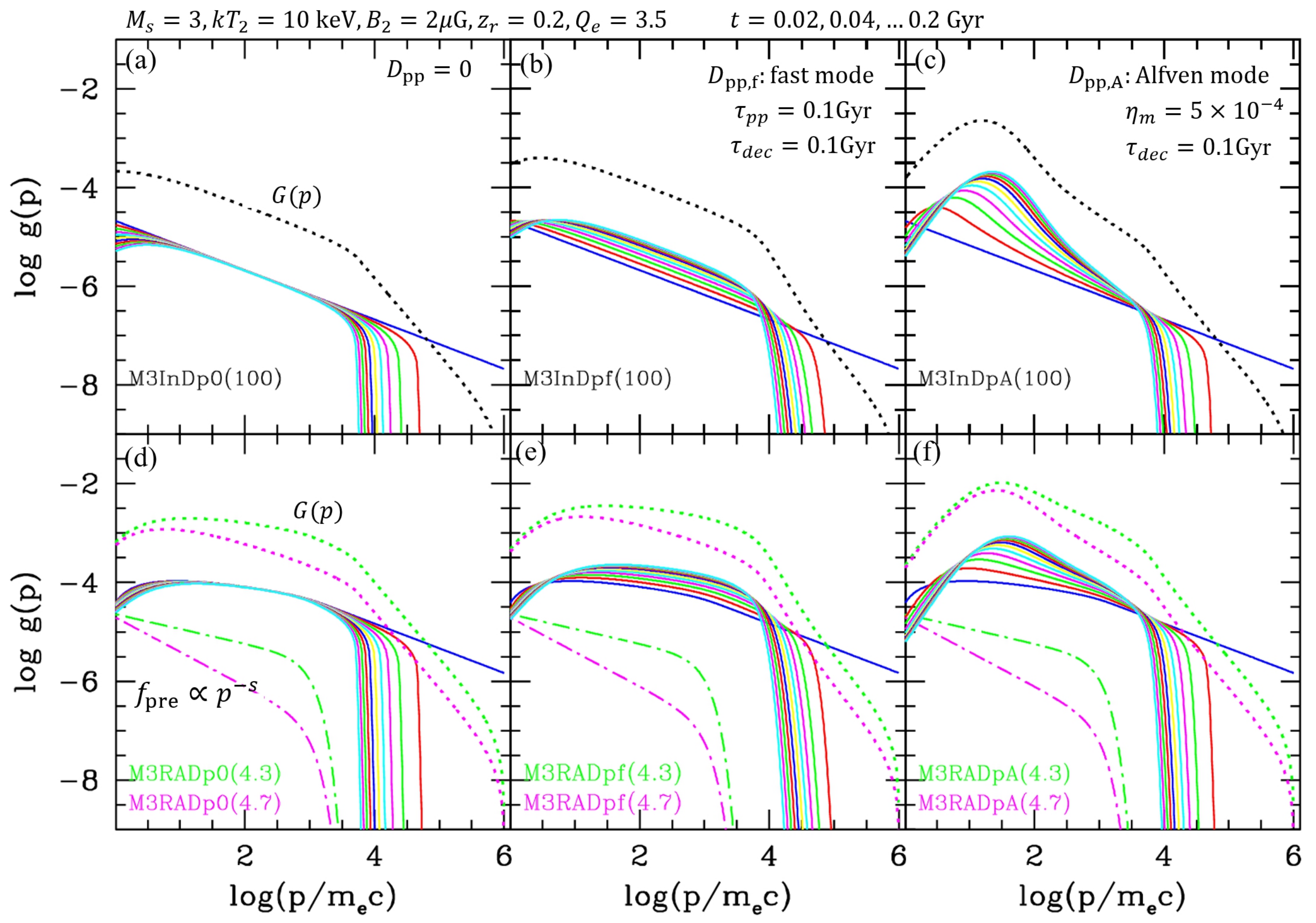}
\caption{Evolution of momentum distribution function, $g(p)=p^4f(p)$, at the avection time, $t=0.02, 0.04, ... 0.2$~Gyr behind the  $M_s=3$ shock models, illustrating the postshock aging with the color coded lines. See Table \ref{t1} for the model names and parameters.
(a-c): The M3In* models are presented. The dotted line in each model represents the volume-integrated spectrum, $G(p)=p^4\cdot u_2\int_0^{t_f} f(p,t) dt$.
(d-f): The M3RA*(4.3) models with $s=4.3$ and $p_{\rm cut}=10^3 m_ec$ are displayed, including the green dotted-dashed line for $f_{\rm pre}(p)$ and the green dotted line for $G(p)$. 
Additionally, for comparison, $f_{\rm pre}(p)$ and $G(p)$ for the M3RA*(4.7) models with $s=4.7$ are shown in the magenta lines.
All functions are given in arbitrary units, but the relative amplitudes among different models are valid. For all models, the decay timescale for postshock magnetic turbulence is set as $t_{\rm dec}=0.1$~Gyr. 
\label{f4}}
\end{figure*}

\subsection{Model Parameters\label{s2.3}}

We choose shocks with Mach numbers $M_s=2.3$ and $M_s=3.0$ as the reference models.
This selection is based on the observation that the Mach number of radio relic shocks detected in the cluster outskirts typically falls in the range of $2\lesssim M_{\rm rad} \lesssim 5$ \citep{wittor2021}. 
Furthermore, numerous particle-in-cell (PIC) simulations have shown that only supercritical shocks with $M_s\gtrsim 2.3$ can effectively accelerate CR electrons in weakly magnetized ICM characterized by $\beta\sim 50-100$ \citep[e.g.,][]{kang2019, ha2021,boula2024}.

The columns 1-4 of Table \ref{t1} list the model names for shocks with $M_s=3.0$, along with the various model parameters being considered.
In M3In* models, the shock-injected population given in Equation (\ref{finj}) is deposited at the shock location, while the reaccelerated population given in Equation (\ref{freacc}) is used in M3RA* models.
Additionally, we will present the same set of models with $M_s=2.3$, denoted as M2.3*, in Section \ref{s3}.

M3InDp0 corresponds to the conventional DSA model without TA ($D_{pp}=0$) in the postshock region with a constant magnetic field ($B_2$).
The effects of decaying $B(t)$ is explored with the two values of the decay time, $t_{\rm dec}=100$~Myr and $200$~Myr.
For M3In* models, the number in the parenthesis represents $t_{\rm dec}$ in units of Myr.
Additionally, we investigate the dependence on the momentum diffusion models, namely $D_{\rm pp,f}$ and $D_{\rm pp,A}$. 
Note that for the models with nonzero $D_{\rm pp}$, the constant $B$ field case is not included, as it is incompatible with the decaying model for magnetic fluctuations.

For M3RA* models, we explore two values of the power-law slope, $s=4.3$ and $4.7$, considering the DSA slope $q=4.5$ for $M_s=3$ shocks.
Note that the number in the parenthesis of the model names for M3RA* represents the value of $s$.

\begin{figure*}[t]
\centering
\includegraphics[width=150mm]{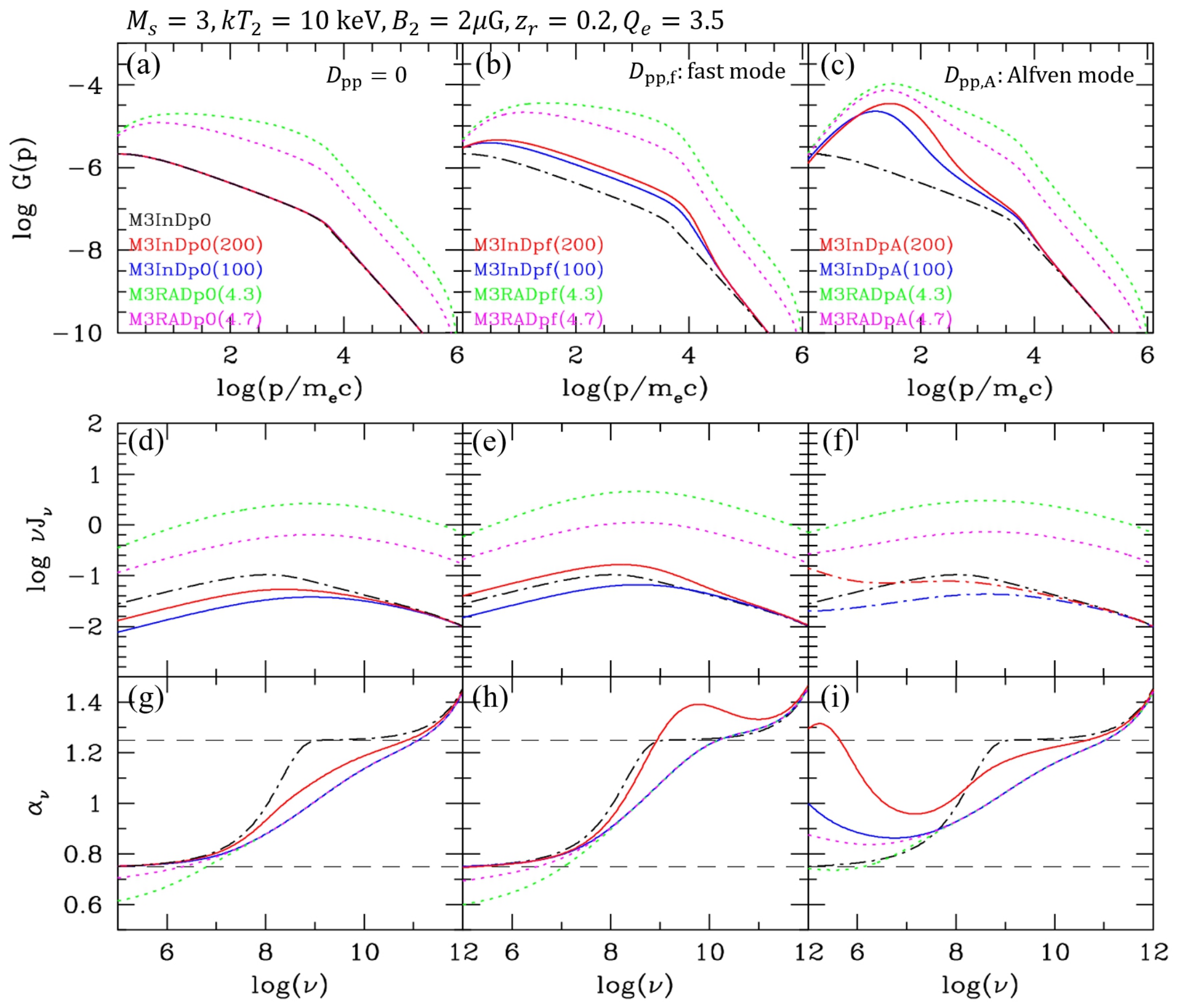}
\caption{(a-c): Volume-integrated spectrum, $G(p),$ for different models with $M_s=3$. See Table \ref{t1} for the model names and parameters. In each column (from top to bottom), the lines and the model names have the same color. 
The M3InDp0 model (no TA and constant $B$) is displayed in the black dotted-dashed line in each panel for comparison.  
(d-f): Volume-integrated radio spectrum, $\nu J_{\nu}$, for the same models shown in the top panels.
(g-i): Spectral index, $\alpha_{\nu}= -d \ln J_{\nu}/d \ln \nu $, for the same models shown in the top panels.
All functions except $\alpha_{\nu}$ are given in arbitrary units, but the relative amplitudes among different models are valid.
For all the models, the total advection time is set as $t_f=0.2$~Gyr.
\label{f5}}
\end{figure*}

\subsection{ Evolution of CR Spectrum in the Postshock Flow\label{s2.4}}
To follow the time evolution of $f(p,t)$ along the Lagrangian fluid element, we solve the following Fokker-Planck equation:
\begin{eqnarray}
{d f(p,t)\over d t} = ({1\over3} \nabla\cdot \mathbf{u}) p {{\partial f}\over {\partial p}} + {1 \over p^2}{\partial \over \partial p} \left[p^2 b_l{\partial f\over {\partial p}} \right] \nonumber\\
 + {1 \over p^2}{\partial \over \partial p}\left[ p^2 D_{pp} {\partial f\over \partial p} \right] 
+ S(p).
\label{FP}
\end{eqnarray}
Here, the cooling rate, $b_l$, includes energy losses from Coulomb, synchrotron, and IC interactions.
Standard formulas for these processes can be found in various previous papers, such as \citet{brunetti2014}. 
Specifically, the Coulomb interaction depends on the density of thermal electrons, $n$, synchrotron losses depend on the magnetic field strength, $B$, and the inverse Compton scattering off the cosmic background radiation depends on the redshift, $z_r$ (see Figure \ref{f2}). 
The divergence term becomes $\nabla\cdot \mathbf{u}=0$ in the postshock flow in 1D plane-parallel geometry, and the source term $S(p)$ accounts for $f_{\rm inj}(p)$ and  $f_{\rm RA}(p)$ deposited at the shock position.

\section{Results\label{s3}}

\subsection{Postshock Cooling and TA of CR Electrons\label{s3.1}}

Figure \ref{f4} illustrates the evolution of the distribution function, $g(p)=p^4f(p)$, for M3In*(100) and M3RA*(4.3) models. Additionally, it presents the volume-integrated spectrum, $G(p)=p^4F(p)= p^4 \cdot u_2\int_0^{t_f} f(p,t) dt$, where $t_f=0.2$~Gyr denotes the final advection time.
The M3InDp0(100) model, which solely incorporates radiative cooling without TA, serves as a reference for comparison with other models.
In Panel (a), it is evident that Coulomb loss is important only for low-energy electrons with $\gamma <10$, whereas synchrotron + IC losses are significant for $\gamma > 10^3$. 
This panel demonstrates that the volume-integrated CR spectrum $F(p)$ steepens from $p^{-q}$ to $p^{-(q+1)}$ above the ``break momentum'' as expected:
\begin{equation}
\frac{p_{\rm br}}{m_ec}  \approx  10^4 \left(\frac{t} { 0.1 {\rm Gyr}}\right)^{-1} \left({B_{\rm e} \over
{5 \mu {\rm G}}}\right)^{-2},
\label{pbr}
\end{equation}
where the effective magnetic field strength, $B_{\rm e}^2= B_2^2 + B_{\rm rad}^2$, takes account for radiative losses due to both synchrotron and IC processes, and $B_{\rm rad}=3.24\mu {\rm G}(1+z_r)^2$ corresponds to the cosmic background radiation at redshift $z_r$.

Figures \ref{f4}(b-c) illustrate how TA with $D_{\rm pp,f}$ or $D_{\rm pp,A}$ delays or reduces the postshock cooling, enhancing $f(p)$. Consequently, the resulting spectrum, including both $f(p,t)$ and $F(p)$, deviates significantly from the simple DSA predictions that take into account only postshock cooling.
As shown in Figure \ref{f2}, TA with $D_{pp,A}$ is dominant for $\gamma<10^2$, while TA with $D_{\rm pp,f}$ becomes more effective for higher $\gamma$ for the parameters considered here.
Regarding the parameter dependence, obviously, TA with $D_{\rm pp,f}$ becomes less efficient for a greater value of $\tau_{\rm Dpf}$.
On the other hand, TA with $D_{pp,A}$ becomes more efficient with a stronger $B$ and a smaller $\eta_m$.

Figures \ref{f4}(d-f) present similar results for the M3RA*(4.3) models, wherein the reaccelerated spectrum $f_{\rm RA}(p)$ with $s=4.3$ is introduced at $t=0$. For illustrative purposes, the normalization factor, $f_{\rm o}$, is set to be the same as that of $f_{\rm inj}(p)$ in Equation (\ref{finj}). Consequently, the resulting $f_{\rm RA}$ (depicted by the blue lines at $t=0$ in the lower panels) is larger than $f_{\rm inj}$ (represented by the blue lines at $t=0$ in the upper panels), as shown in the figure. For the M3RA*(4.7) models with $s=4.7$, only $f_{\rm pre}(p)$ and $G(p)$ are displayed in the magenta lines for comparison. 
In the case of the reacceleration models, both the postshock spectrum, $f(p,t)$, and the volume-integrated spectrum, $F(p)$, may not be represented by simple power-law forms, even without TA.

\begin{figure*}[t]
\centering
\includegraphics[width=150mm]{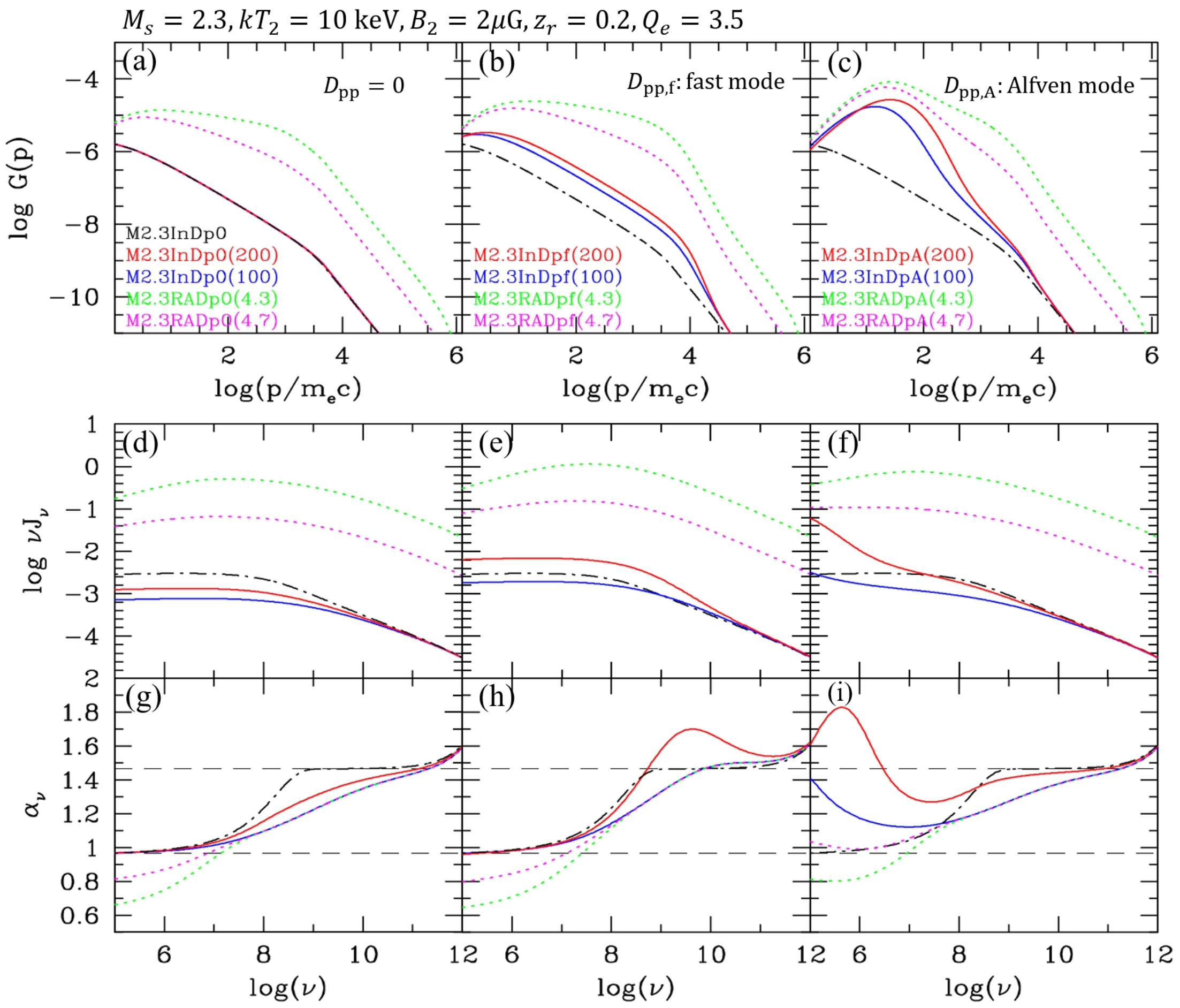}
\caption{The same as Figure \ref{f5} except that $M_s=2.3$ models are shown.
\label{f6}}
\end{figure*}

\subsection{Volume Integrated Radio Emission\label{s3.2}}

\begin{figure*}[t]
\centering
\includegraphics[width=150mm]{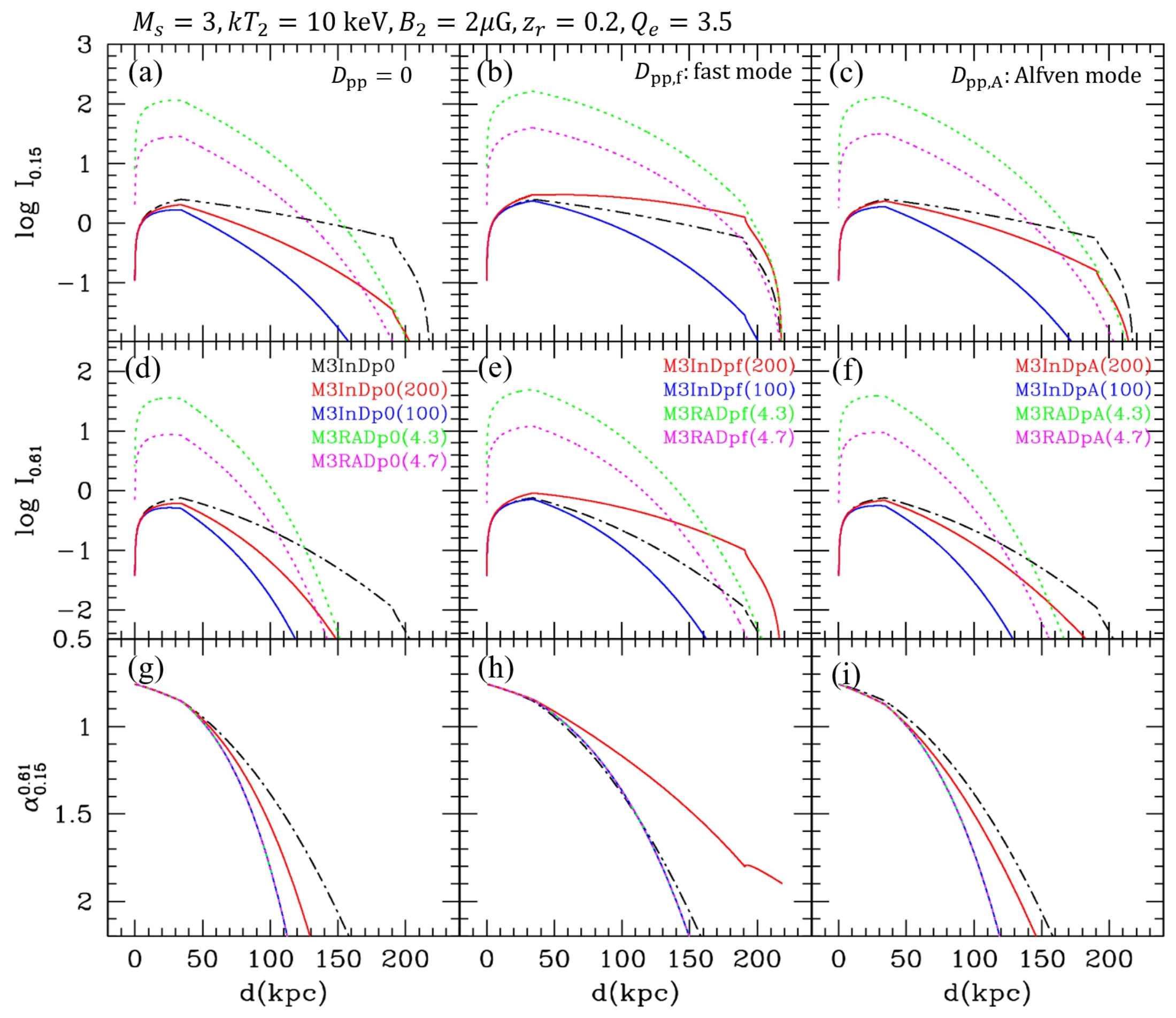}
\caption{(a-c): Surface brightness profile at 0.15~GHz, $I_{0.15}(d)$, for the same M3 models presented in Figure \ref{f4}. See Table \ref{t1} for the model names and parameters. In each column (from top to bottom), the lines and the model names have the same color. In the M3InDp0 model(black dotted–dashed lines), the postshock magnetic field remain constant as $B_2$ and $D_{pp}=0$ (no TA). See Figure \ref{f1} for the adopted shape of the relic surface and the definition of the intensity, $I_{\nu}(d)$. The extension angles are $\psi_1=\psi_2=15^{\circ}$. The displayed functions are given in arbitrary units, but the relative amplitudes among different models are valid.  
(d-f): Surface brightness profile at 0.61~GHz, $I_{0.61}(d)$, for the same models as in (a-c).
(g-i): Spectral index between 0.15 and 0.61~GHz, $\alpha_{0.15}^{0.61}$, for the same models shown in the upper panels.
\label{f7}}
\end{figure*}

Figures \ref{f5}(a-c) compare $G(p)$ for all $M_s=3$ models listed in Table \ref{t1}.
For the three models without TA but with different values of $t_{\rm dec}$, M3InDp0, M3InDp0(200), and M3InDp0(100),
$G(p)$ is almost the same since the total cooling is dominated by the IC cooling, and the effects of decaying $B(t)$ are relatively minor.
For comparison, $G(p)$ for M3InDp0 (no TA and a constant $B_2$) is displayed in the black dotted-dashed line in each panel.
Panels (b) and (c) show the effects of TA with $D_{\rm pp,f}$ and $D_{\rm pp,A}$, respectively.
Thus, compared with the conventional DSA model, TA due to postshock turbulence may enhance the CR electron population.
In addition, the reaccelerated spectrum, $f_{\rm RA}$ (green and magenta dotted lines) could be higher than $f_{\rm inj}$, depending on the amplitude of the fossil electron population.

For the same thirteen models depicted in Figures \ref{f5}(a-c), the volume-integrated synchrotron spectrum, $\nu J_{\nu}$, is shown in Figures \ref{f5}(d-f), while its spectral index, $\alpha_{\nu}= -d \ln J_{\nu}/d \ln \nu $ is displayed in Figures \ref{f5}(g-i).
Again, in each panel, the black dotted-dashed line represents the results for M3InDp0, included for comparison. 
In Panels (d) and (g), the three models without TA, M3InDp0, M3InDp0(200) and M3InDp0(100), are depicted in the black, red, and blue lines, respectively. They demonstrate that the effects of decaying $B(t)$ are quite prominent due the strong dependence of the synchrotron emissivity on the magnetic field strength.
For example, $j_{\nu} \propto B^{(q-1)/2}$ for the power-law spectrum of $f(p)\propto p^{-q}$.

In the conventional DSA model with a constant $B$ (M3InDp0), the transition from $\alpha_{\rm sh}$ to $\alpha_{\rm int}$ occur rather gradually around the break frequency, 
\begin{equation}
\nu_{\rm br}\approx 0.25~{\rm GHz} \left( {t_{\rm age} \over {0.1 {\rm Gyr}}} \right)^{-2}
 \left( {B_{\rm e} \over {5 \mu {\rm G}}} \right)^{-4} \left( {B_2 \over {2 \mu {\rm G}}} \right).
\label{fbr}
\end{equation}
So one should use radio observations at sufficiently high frequencies, $\nu \gg \nu_{\rm br}$, to estimate the Mach number given in Equation (\ref{Mrad2}) using the integrated spectral index \citep{kang15}.
However, as depicted in the red and blue solid lines in Panel (g), this transition takes place much more gradually in the case of decaying magnetic fields with smaller $t_{\rm dec}$. 
Thus, an accurate model for the postshock $B(x)$ is required to estimate the Mach number of radio relic shocks using Equation (\ref{Mrad2}), considering the observational radio frequency range of $\sim 0.1-30$~GHz.

Figures \ref{f5}(h-i) illustrate that TA with a large momentum diffusion coefficient, especially $D_{\rm pp,A}$, could lead to a significant deviation from the simple DSA prediction with a constant magnetic field strength.
We also note that, in Panels (g)-(i), the blue, green, and magenta lines (all with $t_{\rm dec}=100$~Myr) overlap with each other, except for very low frequencies ($\nu < 10$~MHz), whereas they differ significantly from the black ($t_{\rm dec}=\infty$) and red ($t_{\rm dec}=200$~Myr) lines.
This implies that the magnetic field distribution plays a significant role in governing the integrated spectral index $\alpha_{\nu}$ of the volume-integrated radio spectrum.

In Table \ref{t1} for the M3* models, the columns 5-7 list the integrated spectral index, $\alpha_{\nu_1}^{\nu_2}$, between two frequencies, $\nu_1$ and $\nu_2$, where $\nu=0.15,~0.61,~3.0,~{\rm and}~16$~GHz are chosen as representative values.
Moreover, the columns 8-10 list the {\it integrated} Mach number, $M_{\nu_1}^{\nu_2}$, estimated based on Equation (\ref{Mrad2}) using $\alpha_{\nu_1}^{\nu_2}$.   
For M3InDp0, the results are consistent with conventional DSA predictions except for the low frequency case: i.e., $\alpha_{\nu_1}^{\nu_2}=1.25$ and $M_{\nu_1}^{\nu_2}=3$ for $\nu \gg \nu_{\rm br}$.
In the case of $\alpha_{0.15}^{0.61}$, the frequencies are not sufficiently high, resulting in the overestimation of Mach number, $M_{0.15}^{0.61}=3.8$ for M3InDp0.
In fact, for most other models, $\alpha_{0.15}^{0.61}< 1$, so $M_{0.15}^{0.61}$ cannot be estimated.
Both TA and reacceleration significantly influence the integrated spectrum $J_{\nu}$ and tend to generate smaller $\alpha_{\nu_1}^{\nu_2}$, resulting in higher $M_{\nu_1}^{\nu_2}$ except for M3InDpf(200) (see also Figures \ref{f5}(g-i)). 

The M2.3* models also exhibit similar results, as can be seen in Figure \ref{f6}.

\begin{figure*}[t]
\centering
\includegraphics[width=150mm]{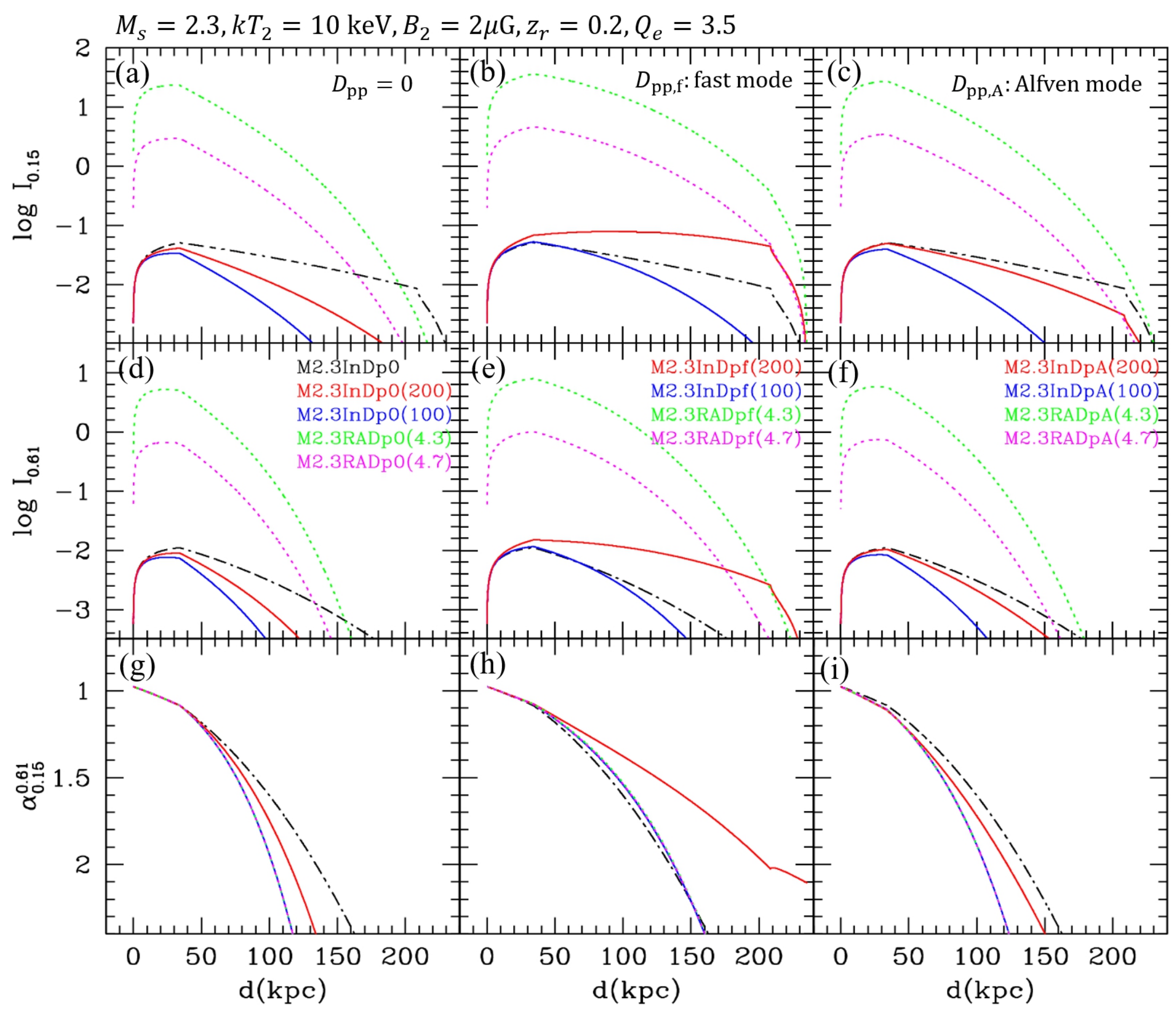}
\caption{The same as Figure \ref{f7} except that $M_s=2.3$ models are shown.
\label{f8}}
\end{figure*}

\subsection{Surface Brightness Profile of Model Radio Relics\label{s3.3}}

Using the geometrical configuration of the shock surface depicted in Figure \ref{f1}, we estimate the surface brightness, $I_{\nu}(d)$, as a function of the projected distance, $d$.
In brief, a radio relic has a coconut-shell-shaped, elongated surface with an axial ratio $a/b\sim 1-1.5$ and a thickness corresponding to the cooling length of electrons, $l_{\rm cool}$.
Here, the radius of the spherical shell is set as $R_s=1$~Mpc.
Then the surface brightness or intensity is calculated by
\begin{equation}
I_{\nu}(d)= \int_{h_{\rm min}}^{h_{\rm max}} j_{\nu}(x) dh,
\label{Inu}
\end{equation}
where $h_{\rm min}$ and $h_{\rm max}$ are determined by the extension angles, $\psi_1$ and $\psi_2$.
As illustrated in Figure \ref{f1}, the path length $h$ along the observer's line of sight reaches its maximum at $d_{\rm peak}=R_s(1-\cos \psi_1)$. 
So for the assumed model parameters, $R_s=1$~Mpc and $\psi_1=\psi_2=15^{\circ}$, the surface brightness peaks at $d_{\rm peak}\approx 34$~kpc.

Figure \ref{f7} presents the spatial profiles of $I_{\rm 0.15}(d)$ at 0.15~GHz and $I_{\rm 0.61}(d)$ at 0.61~GHz for the same thirteen models shown in Figure \ref{f5}.
The spectral index $\alpha_{0.15}^{0.61}(d)$ is calculated from the projected $I_{\nu}(d)$ between the two frequencies. 
Several points are noted:
\begin{enumerate}
\item The postshock magnetic field plays a key role in determining the profile of $I_{\nu}(d)$ and $\alpha_{\nu}(d)$, as it governs the synchrotron emissivity $j_{\nu}$ and $D_{\rm pp,A}$. Consequently, the results depend sensitively on the decay of $B(t)$ in the postshock region.
\item The models with postshock TA (middle and right columns) exhibit a slower decrease in $I_{\nu}(d)$ compared to the models without TA (left column). This occurs because TA delays the postshock cooling of electrons, resulting in a broader effective width of radio relics. In particular, the models with $D_{\rm pp,f}$ generate greater widths than those with $D_{\rm pp,A}$.
\item In the models with $D_{\rm pp,A}$, the enhancement by TA is less significant due to the effects of decaying magnetic fields, distinguishing it from models with $D_{\rm pp,f}$.
\item Panels (g-i) demonstrate that the postshock profile of $\alpha_{\nu}$ is independent of the injection spectrum (i.e., $f_{\rm inj}$ or $f_{\rm RA}$). The profile is mainly influenced by the decay profile of $B(x)$ and by TA due to $D_{\rm pp}(p,x)$. 
\item The spectral index is the smallest at the relic edge ($d=0$), while the intensity profile peaks at $d_{\rm peak}$ in our model set-up for the relic shock surface. Therefore, in observations of radio relics, the region $d<d_{\rm peak}$ corresponds to the postshock region rather than the preshock region.
\end{enumerate}

The M2.3* models presented in Figure \ref{f8} also exhibit the similar behaviors.

\section{Summary \label{s4}}

Giant radio relics are thought to be generated by weak bow shocks that form after the DM core passage during major mergers of galaxy clusters. In such a scenario, CR electrons are accelerated mainly via the Fermi-I mechanism, resulting in the simple predictions for the DSA power-law spectrum, $f(p)\propto p^{-q}$, and the ensuing synchrotron radiation spectrum, $j_{\nu}\propto \nu^{-\alpha_{\rm sh}}$.
Although most observational aspects of radio relics are consistent with such DSA predictions, the so-called Mach number discrepancy among the estimated Mach numbers based on various methods, i.e.,
$M_{\rm rad,int}\gtrsim M_{\rm rad, sh}\gtrsim M_{\rm X}$, remains yet to be resolved.

The ICM is turbulent by nature.
The cascade of magnetic turbulence from large MHD scales to small kinetic scales and the excitation and amplification of magnetic fluctuations via plasma microinstabilities behind the shock front could influence the CR energy spectrum through Fermi-II acceleration.
Moreover, magnetic turbulence is expected to decay approximately in one eddy turnover time, $L/u_2 \sim 0.1$~Gyr, and decaying magnetic fields could significantly affect turbulent acceleration (TA) and the synchrotron emissivity in the postshock region.

In this study, we adopt simplified models for the momentum diffusion coefficient, $D_{\rm pp,f}$ due to fast-mode waves and $D_{\rm pp,A}$ due to Alfv\'en-mode waves, to explore the effects of TA.
The CR spectrum $f_{\rm inj}(p)$ for the shock-injected population or $f_{\rm RA}(p)$ for the shock-reaccelerated population is deposited at the shock front at $t=0$.
Then the time evolution of $f(p,t)$ is calculated along the Lagrangian fluid element in the time-domain. 
The results are mapped onto the spherical shell, whose geometrical configuration is depicted in Figure \ref{f1}, to estimate the surface brightness profile, $I_{\nu}(d)$, as a function of the {\it projected} distance $d$. 

The main results can be summarized as follows:
\begin{enumerate}
\item TA due to $D_{\rm pp,f}$ and $D_{\rm pp,A}$ could delay the postshock aging of CR electrons, leading to a significant deviation from the simple power-law spectrum (Figure \ref{f4}) and a broader spatial width of the surface brightness of radio relics (Figure \ref{f6}).
\item The postshock aging of the CR electron spectrum is insensitive to the decay of magnetic fields since IC cooling dominates over synchrotron cooling (typically $B_{\rm rad} > B$ in the postshock region) (Figures \ref{f5}(a-c) and \ref{f6}(a-c)).
\item The integrated spectral index, $\alpha_{\nu}$, of the volume-integrated radio spectrum sensitively depends on the postshock magnetic field distribution, whereas it is insensitive to the CR spectrum deposited at the shock front.
For instance, the transition from the power-law index $\alpha_{\rm sh}$ to $\alpha_{\rm int}$ occurs more gradually than predicted by the simple DSA model with a constant postshock magnetic field (Figures \ref{f5}(g-i) and \ref{f6}(g-i)). 
Therefore, observational frequencies should be sufficiently high (i.e., $\nu \gg \nu_{\rm br}$) for estimating the Mach number using the integrated spectral index .
\item On the other hand, the synchrotron emissivity scales as $j_{\nu} \propto B^{(q-1)/2}$ and the momentum diffusion coefficient due to Alfv\'en modes, $D_{\rm pp,A}\propto B^2$. This means that the decay of $B$ fields significantly impacts both the surface brightness, $I_{\nu}(d)$, and the spectral index, $\alpha_{\nu_1}^{\nu_2}(d)$ (Figures \ref{f7} and \ref{f8}).
\item The columns 8-10 of Table 1 indicate that, in most models except the MInDp0 model (no TA and constant $B$), the integrated Mach number, $M_{\nu_1}^{\nu_2}$, estimated using the integrated spectral index, $\alpha_{\nu_1}^{\nu_2}$, between two frequencies $\nu_1$ and $n_2$, tends to be higher than the actual shock Mach number. 
\end{enumerate}

This highlights the critical importance of incorporating accurate models for turbulent acceleration arising from postshock turbulence and the impact of decaying magnetic fields when interpreting observations of radio relics. In particular, the shock Mach number estimated using the integrated spectral index may tend to be larger than the actual Mach number. Therefore, a thorough consideration of these factors is essential for a more precise interpretation of radio relic observations.


\acknowledgments
The author thanks the anonymous referee for constructive feedback.
This work was supported by a 2-Year Research Grant of Pusan National University.





\end{document}